\documentclass[aps,prstab,reprint,groupedaddress,showpacs,nofootinbib]{revtex4-1}

\usepackage{graphicx}
\usepackage{amsmath}
\usepackage{amssymb}

\def\SCI#1#2{\hbox{$#1\cdot 10^{#2}$}}

\def\Frot{{f_{\textnormal{rot}}}}
\def\Mn{{m_{\textnormal{n}}}}
\def\Rmin{{r_{\textnormal{min}}}}
\def\nTOF{{t_{\textnormal{n}}}}
\def\toffset{{t_{\textnormal{offset}}}}

\def\DaEq{\quad.}

\begin{document}


\title{Design and test of a compact and high-resolution time-of-flight measurement device for cold neutron beams}


\affiliation{Institut Max von Laue - Paul Langevin, 71 av. des Martyrs, 38000 Grenoble, France}
\author{Damien Roulier}
\email{roulier.damien@gmail.com}
\author{Valery Nesvizhevsky}
\affiliation{Institut Max von Laue - Paul Langevin, 71 av. des Martyrs, 38000 Grenoble, France}
\author{Beno\^it Cl\'ement}
\author{Guilhem Freche}
\author{Guillaume Pignol}
\author{Dominique Rebreyend}
\author{Francis Vezzu}
\affiliation{LPSC Grenoble, Universit\'e Grenoble Alpes, CNRS/IN2P3, 53 av. des Martyrs, 38026 Grenoble Cedex, France}
\author{Stefan Bae\ss{}ler}
\email{baessler@virginia.edu}
\affiliation{Physics Department, University of Virginia, 382 McCormick Road, Charlottesville, VA 22904, U.S.A.}
\affiliation{Oak Ridge National Lab, Bethel Valley Road, Oak Ridge, TN 37831, U.S.A.}
\author{Alexander Strelkov}
\affiliation{Joint Institute for Nuclear Research, RU-141980 Dubna, Russia}



\date{\today}

\begin{abstract}
A time-of-flight device was developed to characterize wavelength distribution and uniformity of a cold neutron beam. This device is very compact -- the distance of flight is $60$~cm -- but achieves very high resolution -- the intrinsic resolution $\Delta \lambda/\lambda=\SCI{2.4}{-3}$ at $\lambda=0.89$~nm. The time-of-flight device is composed of a fixed slit, a disk rotating up to $216$~Hz and a neutron detector with a thin spherical conversion layer with the chopper slit in its focus. The device accepts the complete angular divergence of the initial neutron beam. The efficiency of neutron detection is constant over the detector area. Systematic corrections caused by neutron scattering in air are minimized due to the reduction of the time-of-flight length. Measurements have been performed on the beamline of the GRANIT experiment at ILL (part of the H172 beamline) on level C, and the first order diffraction peak of the crystal monochromator used for the GRANIT beamline was found to be at $\lambda=0.8961(11)$~nm, and having a width of $\sigma=0.0213(13)$~nm. 
\end{abstract}


\maketitle

\section{Introduction}

In fundamental neutron physics, some experiments accept neutrons only within a very small wavelength band. Low energy neutron beams usually provide neutrons in a large wavelength range. The standard solution for removing the neutrons of unwanted wavelength is to use the reflected beam from a crystal monochromator in the neutron beam line that is tailored to the neutron beam and experiment geometry and the desired neutron wavelength. The monochromator crystal reflects neutrons in a given direction only if a Bragg condition is fulfilled. If not, the neutrons are transmitted, potentially for use in other experiments. For a given experiment position, the crystal monochromator is precisely adjusted to reflect the desired wavelength band. Sometimes potential aging of the monochromator crystals has to be monitored. However, once a monochromator is set up on a beamline, it may be difficult to remove because of irradiation of the monochromator and its housing.  Results may be inconclusive if a test beam used has different properties. It may also be unwise to relocate the experiment just to set up a time-of-flight (TOF) device in order to measure a precise wavelength spectrum of the reflected beam. These are the reasons why we have designed and built a very compact device, with a total length of $64$~cm, which is capable of measuring the wavelength spectrum of a cold neutron beam without systematic effects related to the finite angular divergence in the neutron beam or corrections to the neutron scattering in air.
Section~\ref{sec:GRANITBeamline} describes the GRANIT beamline. Section~\ref{sec:dim} describes the specifications and design constraints for our TOF device, and section~\ref{sec:device} describes the device which has been built for our purpose. The obtained results are presented and discussed in section~\ref{sec:resu}.

\section{The GRANIT beamline}
\label{sec:GRANITBeamline}

\begin{figure}
\centering
\includegraphics{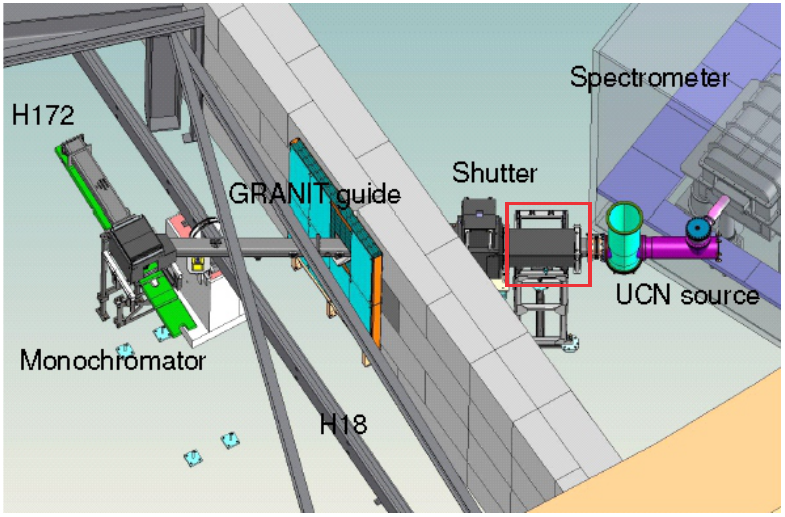}
\caption{Sketch of the GRANIT beamline. Neutrons come from the top left corner through the H172 guide (only its last section is shown). The large box at its end contains the monochromator crystal. The $0.89$~nm neutron beam exits to beam-left, and ends just before the converter volume of the UCN source. For neutron wavelength spectrum  measurements, its last element (marked with a red rectangle in the drawing) has been replaced with the TOF device described in this paper.}
\label{fig:GRANITZone_Setup_Sketch}
\end{figure}

The GRANIT spectrometer is installed on the level C in the reactor building of the Institute Laue Langevin (ILL). The GRANIT instrument is described in \cite{Kre09,CR2010,Pign14}. The spectrometer needs ultracold neutrons (UCN) \cite{Lush69,Stey69}, which are produced in a dedicated UCN source \cite{SchmidtWellenburg2009267}. The purpose of the source is to down-convert cold neutrons to the UCN energy range in liquid helium; this process is most efficient for $0.89$~nm neutrons \cite{Golub77,Yosh92,*Golub92Com,*Yosh92Rep,Baker2003}. The general scheme of the GRANIT beamline is shown in Fig. \ref{fig:GRANITZone_Setup_Sketch}. Cold neutrons with a broad wavelength distribution are delivered from a liquid-deuterium cold neutron source installed in vicinity of the active zone in the ILL high-flux reactor through the H172 neutron guide. The guide geometry and choice of guide walls was optimized to provide a high neutron flux at large neutron wavelengths. The guide has a vertical curvature radius of 650~m upwards, a horizontal curvature radius of 2650~m to beam-left, a length of 14~m, and a square cross-section of 8~cm by 8~cm; the guide walls are coated with $m=2$ Ni-Ti supermirror everywhere but at the bottom where $m=3$ Ni-Ti supermirror was used. Only the last section of this guide is shown in Fig. \ref{fig:GRANITZone_Setup_Sketch}. At its end, there is an intercalated-graphite monochromator (see Ref. \cite{Courtois2011S37}). The monochromator is composed of 18 intercalated stage-2 KC$_{24}$ crystals, with a lattice constant of $d=0.874$~nm. The angle between the white beam and the diffracted beam is $2 \theta = 61.2$~degrees, and the first order peak should be $\lambda = 0.89$~nm. That means, neutrons with a wavelength of $0.89$~nm are reflected towards the UCN source of the GRANIT spectrometer. They are guided through the GRANIT guide (H172-1 to H172-4) with a total length of $4.5$~m, tapered from a cross section of 8~cm by~8 cm to 7~cm by 7~cm, in order to optimize neutron flux in the UCN converter volume (which at that time was to be designed). The GRANIT guide walls are coated with $m=2$ Ni-Ti super-mirror. For our measurements, the last section of the tapered guide (H172-4, length $1.1$~m) has been taken out. The $^4$He UCN source is installed downstream of a neutron shutter built into this guide, and is described in Refs. \cite{Zimm11, Pieg14}. UCN are produced in the source and delivered to the GRANIT spectrometer installed inside a clean room using a dedicated UCN extraction system. Its design is described in \cite{SchmW07Extract}. An update, and characterizations of the various beam line elements are given in \cite{Roul15}. The motivation for this work was to understand the spectral neutron flux, which may help to better understand the GRANIT beamline. In particular, if the mean wavelength in the $0.89$~nm peak would considerably differ from the desired value, we would be obliged to reinstall the GRANIT neutron guides. One should note that the choice of output angle of the crystal monochromator is not a simple geometrical problem because the H172 guide is curved, which causes the neutron beam direction to be different from a tangent to the end of the H172 guide.

\section{Design constraints}
\label{sec:dim}

A usual method to determine the wave length spectrum of a neutron beam is to measure the time of flight (TOF) of the neutrons. For this work, the goal is to use a compact device with an intrinsic wavelength resolution of about 0.1\% -- the relative spectral width of the $0.89$~nm peak in the wavelength distribution is about 1\%. The device should be movable as one piece as the acceptance area is small, and the device should allow moving it over the cross section of the beam. Furthermore, the device needs to be short to fit between the GRANIT shutter and the UCN source (see Fig. \ref{fig:GRANITZone_Setup_Sketch}, the measurement device will replace the section of the GRANIT guide downstream of the shutter). The detection efficiency for neutrons is desired to be proportional to wavelength for all neutrons between $0.2$~nm and $1.0$~nm, so as to measure the neutron capture flux undistorted. Our TOF device consists of a rotating thin chopper disc behind a entrance collimator, and a neutron detector mounted in a certain distance behind it. The chopper disc acts as a gate that opens the neutron beam for a short time. The opening of this gate starts a clock. When a neutron arrives in the detector at the downstream end of the TOF device, the clock is read out. For a length of the neutron flight path $D$ between chopper disc and detector, the time of flight of a neutron $\nTOF$ is related to the neutron wavelength through
\begin{equation}
\nTOF=\frac{D \Mn}{h} \lambda \DaEq
\label{eq:TOF}
\end{equation}
The TOF of the neutrons $\nTOF$ is experimentally determined through recording the time difference between a ``start'' signal -- which occurs when the entrance slit before the chopper is aligned with one of the slits in the chopper disc -- and a ``stop'' signal -- the detection of the neutron.
We refer to the introduction in Ref. \cite{Abe06} for a thorough description of potential pitfalls of the TOF method, related to detector size, type, position, extra collimation, and requirements on the data acquisition system. 
The reconstruction of the wavelength through eq. \eqref{eq:TOF} can be easily made precise by determining very precisely the flight path $D$ -- this is a simple mechanical problem -- and $\nTOF$ -- this is a simple electronics problem. Often, the calibration of the time of flight scale is performed through a comparison of measurements performed with the detector installed at different distances from the slit in the chopper; here we used a different method that is described below. 

As we wanted to avoid corrections for the efficiency of the TOF device as a function of neutron wavelength, we had to design and build a detector with a size that is sufficient to cover the full angular divergence of the neutron beam. In addition, we desire the detector to have uniform efficiency over the detector surface. Equal flight path length at the desired level of accuracy required the detector surface to be spherical with its focus at the chopper slit. For the neutrons with the largest wavelength of interest, $\lambda_{\textnormal{max}}=1.0$~nm, the beam divergence for a beam from a $m=2$ supermirror\footnote{The $m=2$ stands for a reflectivity of the supermirror that is large for an angle to the surface below $m=2$ times the critical angle for a reflecting surface made from natural Ni. For Ni, neutrons of wavelength $\lambda$ are totally reflected up to an angle of $1.73~\hbox{mrad}/\hbox{\AA} \cdot \lambda$.}
is $\theta=m\cdot 1.73\ \hbox{mrad}/\hbox{\AA} \cdot \lambda_{\textnormal{max}} = 35~\hbox{mrad}$. Taking into account the fact that the guide is slightly converging, the maximum angle neutrons with wavelength below $\lambda_{\textnormal{max}}=1.0$~nm could have is $\theta_{\textnormal{max}}=37.5$~mrad\footnote{The GRANIT neutron guide is long enough that its reflectivity sets the limit for the beam divergence, and not the mosaicity of the crystal. }. The detector that was used had a diameter of $D_{\textnormal{det}}=80$~mm, which means that the maximum displacement from the beam axis in either horizontal or vertical direction for a neutron hitting the detector is $x_{\textnormal{max}}=D_{\textnormal{det}}/2\sqrt{2}\approx 2.8$~cm. This condition requires $D< x_{\textnormal{max}}/\tan(\theta_{\textnormal{max}})=75$~cm, with the exact number depending on the size of the slit in the chopper and entrance collimator. 

The intrinsic resolution of the TOF device, that is, the width of the measured wavelength spectrum for a monochromatic incoming neutron beam, critically depends on the choice of materials and dimensions: The chopper disk should be as thin as feasible. A thick chopper removes neutrons preferably if they have a large divergence as they may enter the absorbing material of the chopper disk from the inside walls of the chopper slit. These are the neutrons with a larger wavelength. Therefore, a chopper deforms the wavelength spectrum due to its thickness. This conditions disqualifies a chopper disk made from boron-loaded aluminum. Furthermore, the thinner chopper disk allows for a more precise definition of the flight path.
One would like the disk to be having a large diameter, and to rotate with high frequency, in order to minimize the time $\Delta t_{\textnormal{gate}}$ during which chopper slit is aligned with the entrance collimator so that the neutron beam can enter the flight path. A compromise must be found in order to keep the TOF device compact, and to find a suitable drive motor for the chopper disc having sufficient torque. The duration of the beam pulse limits the wavelength resolution of the device. The chopper may contain more than one slit; the demand is that the recorded spectra to adjacent slits do not overlap.

The signal-to-background ratio $S/B$ is limited by the integrated transmission of the chopper when the slits are not aligned. The value of $S/B$ is clearly increased by having a thick chopper. However, this is in conflict with the constraints on the choice of the motor, and with the condition of having an unbiased angular acceptance of the beam. The chopper should be made from efficient neutron absorbers; i.e. cadmium or gadolinium are the good candidate materials. Ideally, the chopper slit has the shape of a sector of a hollow circle; for ease of manufacturing, we approximated this by giving the chopper slits a shape of  a rounded rectangle.


\begin{figure*}
\centering
\includegraphics{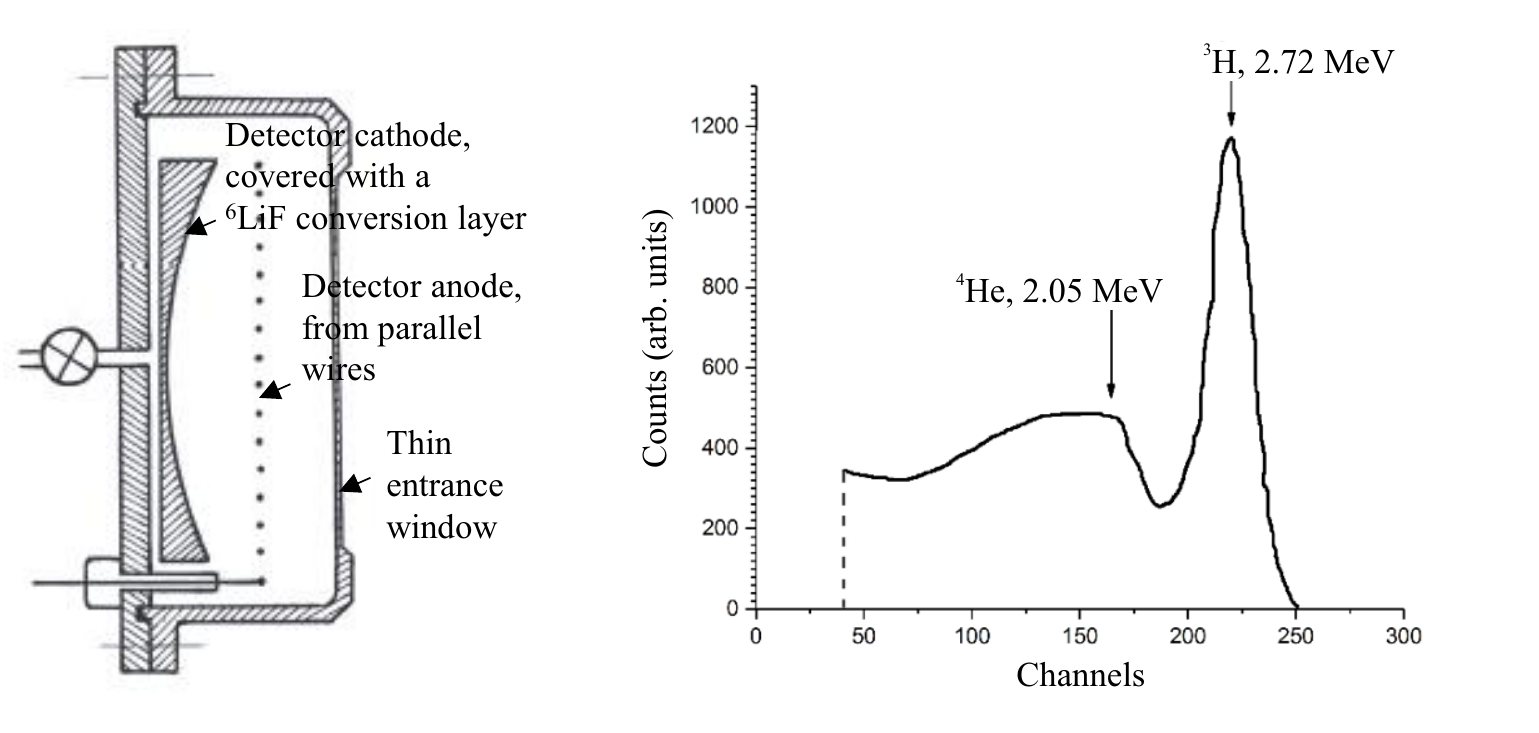}
\caption{Left: Scheme of the detector designed for the TOF device (side view).
Right: Amplitude spectrum of the detector output signal (in a.u.). The two peaks correspond to the detection of $^4$He ($2.05$~MeV) and $^3$H ($2.72$~MeV) from neutron capture in $^6$Li. The lowest amplitudes are cut by a discriminator.
}
\label{fig:newtof_det}
\end{figure*}

\section{The time-of-flight measurement device}
\label{sec:device}
\subsection{Neutron detector with a thin spherical LiF converter}

The detector setup is shown in Fig. \ref{fig:newtof_det}. The detector counts cold neutrons coming through an entrance window with $80$~mm diameter. The radius of curvature of the converter layer, $0.5$~m, is made similar to the distance of flight $D$ to have an almost equal flight path from the chopper to the detector for all neutrons irrespective of their angle to the beam axis. The converter is a $^6$LiF layer with the thickness of $0.54$~mg/cm$^2$. In the $^6$LiF, the Li isotope has 90\% enrichment. The efficiency of conversion, $\epsilon$, on the thin $^6$LiF layer is a few percents for neutrons with $\lambda_{\textnormal{max}}=1.0$~nm. The dependence of $\epsilon$ on the neutron wavelength $\lambda$ ($\epsilon=\epsilon_{0.89\ \textnormal{nm}}\cdot \lambda/0.89\ \textnormal{nm}$) must be taken into account for the spectrum normalization. The LiF layer is applied by thermal evaporation in vacuum onto an Al disk. This disk, with a diameter of 10~cm, is the cathode of a multi-wire proportional counter filled with a gas mixture of Ar and CH$_4$; the partial pressure of CH$_4$ is $\sim 30$~torr; the total gas pressure is 2~bar. The drift velocity of free electrons, which are released from gas ionization by $\alpha$-particles and tritons in the nuclear reaction $^6$Li(n,$\alpha$)$^3$H, is $\sim 1$~cm/$\mu$s in such gaseous mixture. The drift does not result in significant time delays or scattering of times of arrival of signal pulses, which is essential for applying the time-of-flight method used to measure neutron spectra. The detector anode consists of parallel wires with the diameter of $20$~$\mu$m; the distance between neighbor wires is 5~mm; the distance between the wires and the cathode is $1.5$~cm. The electrical capacity of the detector with a short, 20~cm, cable is $\sim 35$~pF. The detector operates in the proportional regime when the applied voltage is $\sim 2$~kV; then the signal amplitude is $\sim 2$~mV and the duration is $\sim 0.5$~$\mu$s. In order to calibrate the neutron detection efficiency, we added temporally a known quantity of $^3$He into the gas mixture in the detector and directly compared the number of counts from the LiF layer and the $^3$He admixture in the amplitude spectrum. For the neutron velocity of 450~m/s, the detection efficiency is $\sim 2.5$\%. The efficiency would not change if the gas pressure decreases down to $0.1$~bar. However, such a decrease significantly decreases the signal duration and the detector voltage. This is possible only in the proportional regime with its gas enhancement for the registration of only a fraction of energy of $^4$He and $^3$H deposited in the gas at low pressure.    
The variation of the efficiency of neutron detection with the position of neutron absorption in the converter layer was measured with a source of thermal neutrons, and was found to be below 5\%.

\subsection{Chopper}

\begin{figure*}
\centering
\includegraphics{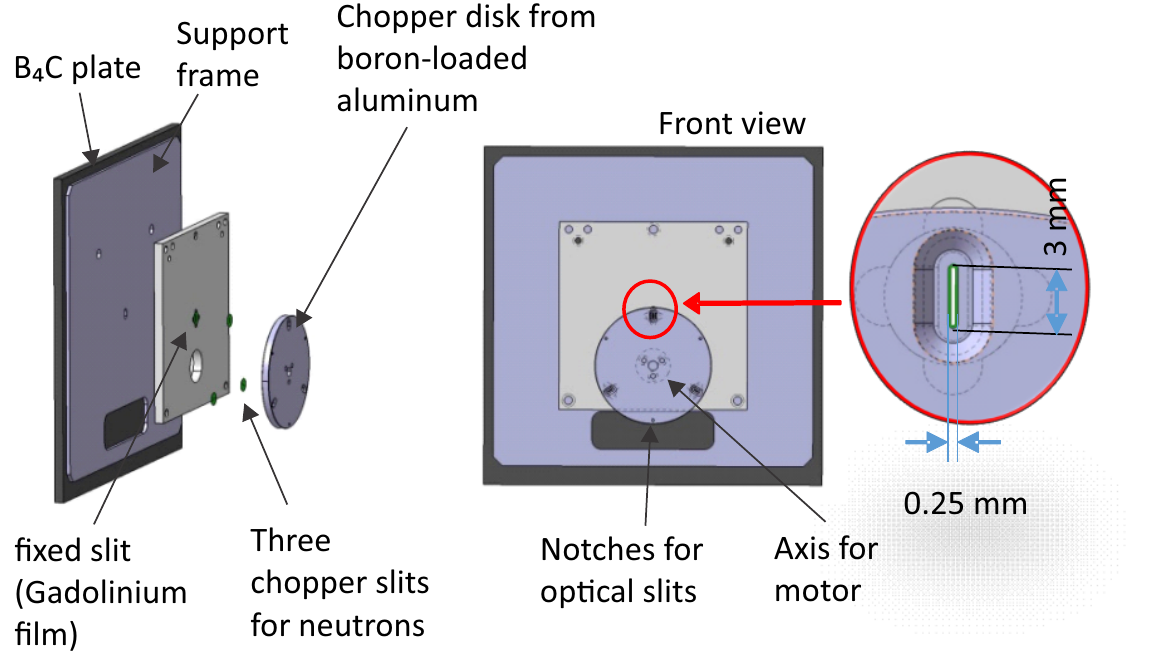}
\caption{The beam chopper, composed of a rotating disk with 3 slits and one fixed slit. The borated aluminum around the gadolinium slits is beveled giving a sufficient aperture angle for the beam when the disk is rotating. Diametrically opposed of each slit is an optical slit at large radius to activate an optical switch. Neutrons cannot pass the optical slit.}
\label{fig:tof_disk}
\end{figure*}

The chopper is shown in Fig. \ref{fig:tof_disk}. The chopper disk contains $n_{\textnormal{slits}}=3$ rectangular slits, each of them with a width of $l=0.25$~mm and a height of $L=3$~mm. The distance between the rotation axis and the inner edge of the chopper slit is $\Rmin=30$~mm. The chopper is designed to rotate with a frequency of $\Frot=167$~Hz.

The chopper disk is made of borated aluminum ($4.5\%$ weight of boron, enriched in $^{10}$B at $95\%$) of width $4$~mm. All slits (including the fixed slit) have a stadium shape, and are made from a $0.1$~mm thick and $1.5$~cm diameter gadolinium foil, for reasons explained above. This design results in a chopper disk with a sufficiently low moment of inertia for the motor chosen, and limits the machining of Gd, a hazardous material. 

The fixed slit consists of a large sheet of 5~mm thick B$_4$C plastic sheet backed by an aluminum plate for mechanical stability. The plastic sheet contains a cutout, in which the same kind of gadolinium foil is mounted, but now with a slit width of $l_{\rm f}=0.25$~mm~ and slit height $h_{\rm f}=3$~mm. The fixed slit and the disk are separated by $0.3$~mm, thus the neutron collimation occurs within $0.5$~mm, and neutrons are very well absorbed around the slits.

The time between two adjacent neutron beam pulses that appear when the fixed slit is aligned with one of the slits on the chopper disk is $T=1/n_{\textnormal{slits}}\Frot=2$~ms. The detector is intended to be placed at a distance of $D\sim60$~cm from the chopper. Thus, the slowest neutrons from the beam that are detected irrespective of their angle relative to the beam axis have a velocity of $300$~m/s, corresponding to a wavelength of $1.32$~nm. Frame overlap does not cause a significant distortion of the spectrum.

\begin{figure*}
\centering
\includegraphics{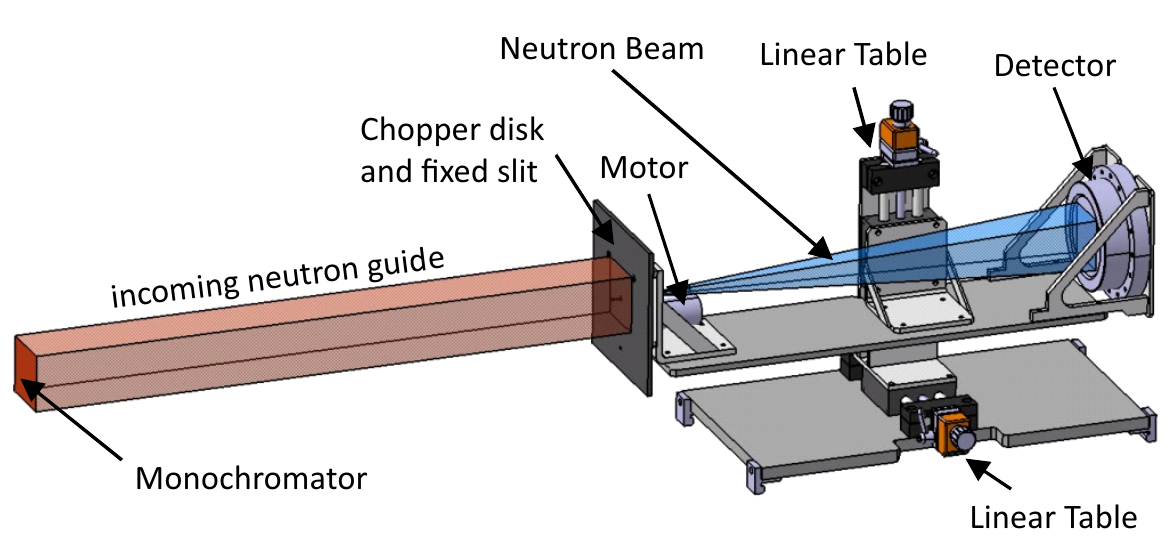}
\caption{The whole TOF measurement device. The fixed slit, the rotation disk and the detector are jointly set on a translation table with two degrees of freedom.}
\label{fig:tof_full}
\end{figure*}

The detector and the chopper are set up rigidly on a 2D translation table, allowing measurements in the orthogonal plane of the beam. This is illustrated in Fig.~\ref{fig:tof_full}. 

The distance of flight is set to $D=60.679\pm0.072$~cm, and the detector covers a solid angle of $0.27$~msr, which is within the specification mentioned above. The uncertainty on the distance of flight $D$ gives a systematic error of $0.12\%$ on the wavelength from eq. \eqref{eq:TOF}.
An upper limit on the instrumental resolution of the device, slightly larger than what it really is, is obtained by assuming a slightly enlarged slits with a shape of a hollow sector of a circle with center at the rotation axis just large enough to contain the true rectangular slit. If so, the chopper is open for all neutrons that pass the slit for the same time, independent of the radial distance of the neutrons to the chopper axis. We compute the intrinsic chopper resolution to be
\begin{equation}
\sigma_a\approx\frac{1}{\sqrt{24}}\cdot \frac{h}{\Mn D}\cdot \frac{l+l_{\rm f}}{2\pi \Rmin\Frot}=0.0021(1)~\hbox{nm} \DaEq 
\end{equation}
The first factor gives the standard deviation of a symmetric triangular distribution of unit width, the second translates a width in time into a width in wavelength, and the last factor is the time for the chopper to be open. The uncertainty $\sigma_a$ is dominated by the uncertainty in $l$ and $l_{\rm f}$. 

\subsection{Electronics}
Data acquisition is performed by the UCTM2 board designed at LPSC~\cite{UCTM2}. This module allows for a variety of ways to treat analogue input signals. The eight input channels are processed through an constant amplitude discriminator for either a positive or a negative signal and 
produces pulse of configurable length and with configurable delay. Trigger signals can then be generated by any logical combination of these windows allowing to trigger on coincidences, anti-coincidences, or more complicated combinations of inputs.
The module then uses these trigger signals for either analog to digital conversion (voltage or charge sensitive) or time to digital conversion. This last feature is used in this experiment.
The sampling resolution is $5$~ns, which means the uncertainty on the time of flight due to the board is insignificant -- the time of flight 
of neutrons with a wavelength of $0.13$~nm is about $2$~ms with our device. 
The ``start'' signal is given by an optical switch which is closed when one of the optical slits located on the periphery of the rotating disk is 
passing the switch, signaling that one of the neutron slits is aligned with the fixed slit, as shown in Fig. \ref{fig:tof_disk}. 
The output signal of the neutron detector is pre-amplified and provides a ``stop'' signal. The board is able to register several ``stop'' signals for each ``start'' signal. 
Two modes are thus available, ``monostop'' and ``multistop''. The second one is used here, as multiple neutrons can be detected during the time a slit is open, and in the ``monostop'' mode, a faster neutron would mask a slower one. 

Due to the positioning of the optical switch, the optical ``start'' signal has a $3.3$~deg misalignment with the slit opening. At a rotation speed of $\Frot=\pm 167$~Hz, it generates a time offset of $\toffset\approx\mp 58~\mu$s (the width of light source and slit opening gives a small correction). This was verified by changing the sense of orientation of the rotation, and by requiring the position of the higher order Bragg peaks at $\lambda/2$ and at $\lambda/3$. 

\section{Tests and results}
\label{sec:resu}


Measurements have been performed on the cold neutron beam of the GRANIT beamline. From previous measurements~\cite{Pieg14,Roul15}, several peaks from higher order in Bragg reflection, or crystal irregularities are expected. The wavelength of interest for the experiment is $\lambda=0.89$~nm, and the previous results were inconclusive about the proportion of neutron beam flux in the diffracted peak in this bandwidth.



Our TOF measurement device has an efficiency for all neutrons within the given wavelength range that is proportional to the neutron wavelength $\lambda$ due to the efficiency of the neutron detector; otherwise than for this detector effect, all neutrons that pass through the fixed slit are equally likely to be detected. Normalization of the TOF spectrum can be done most easily with a gold foil activation measurement.

\begin{figure}
\centering
\includegraphics{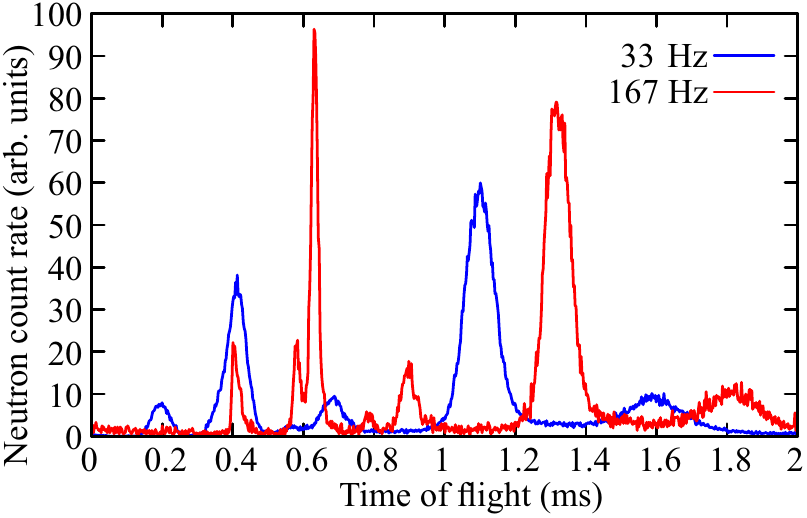}
\caption{TOF spectra obtained with two rotation speeds ($\Frot=33$~Hz and $\Frot=167$~Hz). The times shown are as measured, that is, $\toffset$ has not been corrected for.}
\label{fig:multistop}
\end{figure}

The TOF spectra obtained are shown in Fig.~\ref{fig:multistop}, both at nominal and reduced rotation speed. The expected signal over noise ratio obeys $S/B\geq50$, as it is expected from the dimensions and material of chopper disk and fixed slit. The signal-to-noise ratio would be worse if boron-loaded plastic was used.
The measurement also directly proves that the instrumental resolution scales with the time of flight for a given peak, consistent with the instrumental resolution given above. 

\begin{figure}
\centering
\includegraphics{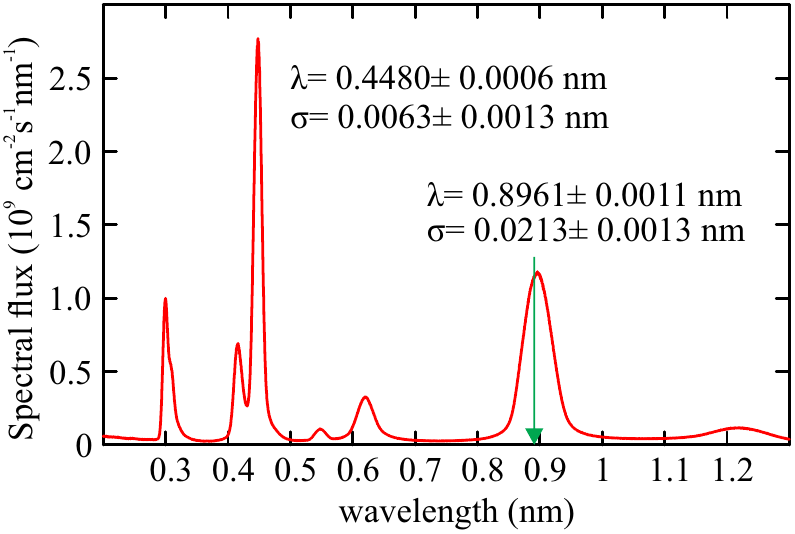}
\caption{Spectral neutron flux (not: neutron capture flux) from the center of the GRANIT beamline, obtained by time-of-flight measurement. The neutron capture flux measured at the position of the device is used for the normalization. The green arrow shows the useful wavelength for UCN production in the LHe source.}
\label{fig:tof_full_spectrum}
\end{figure}

We observe in Fig.~\ref{fig:multistop} that with a lower rotation frequency, the offset generated by the switch misalignment is greater, and the peaks are wider, as expected.
To adjust the total time offset in the acquisition, we use the proportionality of time of flight between the first, second and third orders diffraction peaks. The adjusted value is $\toffset=-56\pm1~\mu$s for nominal rotation speed, close to the value found from the investigations mentioned above. Once this offset is taken into account, the spectrum is converted in wavelength using eq. \eqref{eq:TOF}, and the wavelength-dependent efficiency $\epsilon$ of the detector is corrected for. Neutron absorption in air is neglected in this analysis. For normalization, we have measured the capture flux with gold foil activation. Scaled to a reactor thermal power of $58.2$~MW, the result was $
\Phi_{\textnormal{capture}}=(7.94\pm 0.29)\times 10^8\text{ cm}^{-2}\text{s}^{-1}$ .
We obtain the normalized wavelength spectrum shown in Fig.~\ref{fig:tof_full_spectrum}. This spectrum is taken from the center of the exit window of the GRANIT neutron guide. The centroid of the main diffraction peak is at $\lambda=0.8961\pm0.0011$~nm, while its standard deviation is $\sigma=0.0213\pm0.0013$~nm. The spectral flux at this wavelength in the center is $\left. d\Phi/d\lambda\right|_{0.89~\textnormal{nm}}=1.20(4)\cdot 10^9$~cm$^{-2}$s$^{-1}$nm$^{-1}$ at a reactor power of $58.2$~MW, slightly higher, but within the uncertainties consistent with earlier measurements in Ref. \cite{Pieg14}.

\begin{figure*}
\centering
\includegraphics[width=0.9\textwidth]{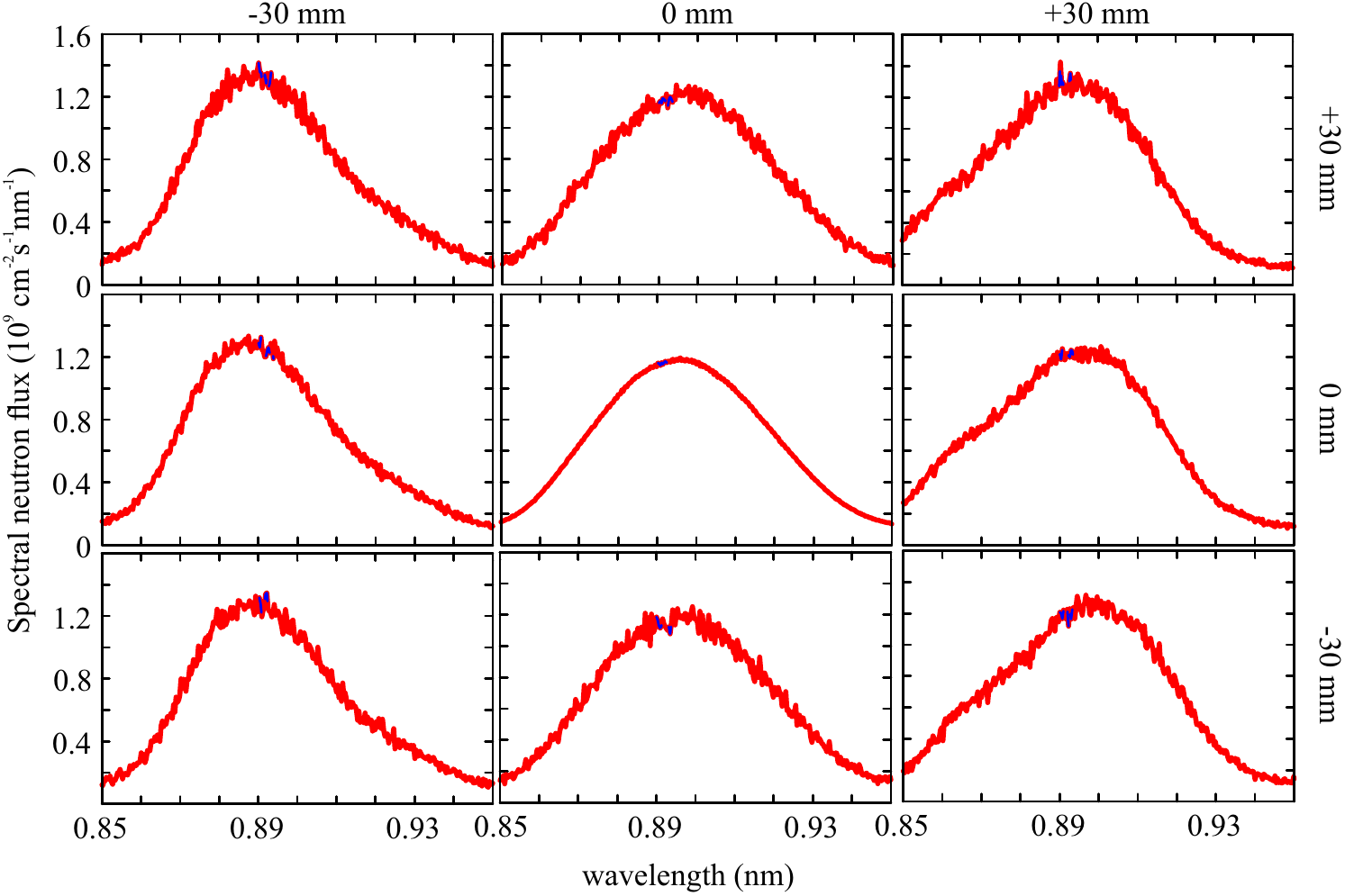}
\caption{Spectral neutron flux obtained by time-of-flight measurement at various positions of the beam. The positions are given relative to the center of the exit window of the GRANIT neutron guide, after the shutter. The blue lines cover the range of interest for UCN production.}
\label{fig:scanbeam}
\end{figure*}

Furthermore, we find that the spectral flux does not depend substantially on the position at the exit window of the neutron beam, as shown in Fig. \ref{fig:scanbeam}. The precision and the resolution of this measurements are now sufficient to conclude that the effective spectral flux at $0.89$~nm is only a few percent lower than what could be obtained with optimum monochromator position. Sub-structures in the spectrum, hidden by the lack of resolution in the previous measurements, have been revealed: The main diffraction peak for a stage-3 graphite monochromator crystal at $\lambda=0.89$~nm is accompanied by higher order peaks at $\lambda/2$ and $\lambda/3$. In addition, we see contaminations consistent with stage-1 KC$_8$ and stage-3 KC$_{48}$ compounds. Our results may be compared with the one obtained for a similar setup at the Fundamental Neutron Physics Beamline at the Spallation Neutron Source (see Ref. \cite{Fomin2015,*Fomin2015C})

Future applications of the new device at ILL are the improvement of the precision of wavelength calibration at D17 instrument \cite{Cubitt02}, which is needed in particular for advanced measurements of neutron whispering gallery modes \cite{Valery2010} associated with searches for extra short-range forces \cite{SR2010}. The TOF device will also be used at SUPER-ADAM instrument \cite{SuperADAM2013} to characterize the neutron beam after the intercalated graphite monochromator, and at the PF1B instrument \cite{Abe06} for measurements in the restricted geometry typical for "user" experiments in nuclear and particle physics. Furthermore, the device will be used on the CT2 beamline (development of neutron detectors) with an intercalated graphite monochromator, where the relative intensities of the diffraction peaks are of interest to evaluate the efficiency of detectors. In the context of the upgrade of the H24 distribution line, several other ILL instruments would also benefit from a re-characterization of their incoming neutron beam.
\vspace{3mm}

\section{Conclusion}

We have designed and built a new time-of-flight device, whose main characteristics are compactness, absence of the need for significant systematic corrections, and a high wavelength resolution to measure the wavelength spectrum of a cold neutron beam in spatially limited environments. This system has been successfully tested at the H172/GRANIT beamline on level C at the ILL, leading to a better  -- and required -- understanding of the UCN production chain for the GRANIT experiment. Thanks to its high resolution, new sub-structures in the beam wavelength spectrum have been revealed and will be studied in order to check the aging of the monochromator.


\begin{acknowledgments}
We thank Michael Kreuz for information about the beam line, and all members of the GRANIT collaboration for their continued support and interest in this work. S.B. acknowledges support from ILL for a scientific visit during which a substantial part of the paper was written.
\end{acknowledgments}


\end{document}